\documentclass[twoside]{article}
\usepackage{fleqn,espcrc2}
\usepackage{colordvi}

\usepackage{epsfig}
\usepackage[figuresright]{rotating}

\newcommand{\cp}{CP}
\newcommand{\pt}{$p^{}_T$}
\newcommand{\ddbar}{$D^0$-$\overline{D}^{\,0}$}

\newcommand{\AmS}{{\protect\the\textfont2
  A\kern-.1667em\lower.5ex\hbox{M}\kern-.125emS}}

\title{
\vspace*{-0.40in}
\begin{flushright}
{\large UCTP-112-00}
\end{flushright}
\vspace*{-0.10in}
CP Violation and Mixing Results from FNAL E791}

\author{A. J. Schwartz\address{
Department of Physics,
University of Cincinnati,
Cincinnati, Ohio 45221} \\
(Representing the E791 Collaboration)}
       
\begin{document}

\begin{abstract}
We review results from FNAL E791 concerning \ddbar\ mixing and \cp\ violation
in $D$ meson decays. We have searched for mixing in semileptonic
$D^0\rightarrow K^+\ell^-\bar{\nu}$ decays and in hadronic
$D^0\rightarrow K^+\pi^-$ and $D^0\rightarrow K^+\pi^-\pi^+\pi^-$ decays. 
We have searched for \cp\ violation in $D^0\rightarrow K^+K^-/\pi^+\pi^-$ and 
$D^+\rightarrow \phi\,\pi^+/\overline{K}^{*0}K^+/K^+K^-\pi^+/\pi^+\pi^-\pi^+$ 
decays. Finally, we have measured the difference in decay widths $\Delta\Gamma$ 
between the two mass-eigenstates of the \ddbar\ system. This parameter affects 
the rate of \ddbar\ mixing. We combine our results with those from other 
experiments to obtain confidence intervals incorporating all published
experimental data. 
\vspace{1pc}
\end{abstract}

\maketitle

\section{INTRODUCTION}

FNAL E791\footnote{The collaboration consists of:
CBPF (Brazil), Tel Aviv, CINVESTAV (Mexico), Puebla (Mexico),
U.~C.\ Santa Cruz, Cincinnati, Fermilab, Illinois Institute of 
Technology, Kansas State, Massachusetts, Mississippi, Princeton, 
South Carolina, Stanford, Tufts, Wisconsin, and Yale.} 
is a hadroproduction experiment studying the
weak decays of charm mesons and baryons. The charm particles were produced
by impinging a 500~GeV/$c$ $\pi^-$ beam on five thin target foils. 
The most upstream foil consisted of platinum; the other foils consisted
of carbon. All foils were separated by about 15~mm such that $D$ mesons 
decayed predominately in the air gaps between foils. The experiment 
took data from September, 1991 to January 1992, recording the world's 
largest sample of charm decays at that time. The number of reconstructed
events is over 200\,000. With this data sample the experiment has studied 
charm production \cite{e791:production,e791:epj}, charm lifetimes 
\cite{e791:lifetimes,e791:gamma_diff}, rare and 
forbidden $D$ decays \cite{e791:rare}, 
\ddbar\ mixing \cite{e791:mixing}, 
\cp\ violation \cite{e791:cp0,e791:cp+}, and several other topics. 
Here we focus on the following:
{\it (a)}\ searches for \ddbar\ mixing in 
semileptonic $D^0\rightarrow K^+\ell^-\bar{\nu}$
decays and in hadronic $D^0\rightarrow K^+\pi^-$ and 
$D^0\rightarrow K^+\pi^-\pi^+\pi^-$ decays;
{\it (b)}\ measurement of the doubly-Cabibbo-suppressed decay
$D^+\rightarrow K^+\pi^-\pi^+$;
{\it (c)}\ search for \cp\ violation in neutral 
$D^0\rightarrow K^+K^-/\pi^+\pi^-$ decays and in charged 
$D^+\rightarrow \phi\,\pi^+/\overline{K}^{*0}K^+/K^+K^-\pi^+/\pi^+\pi^-\pi^+$ 
decays; 
{\it (d)}\ measurement of the width difference $\Delta\Gamma$ between the
two mass-eigenstates of the \ddbar\ system. These results are published 
in Refs.~\cite{e791:gamma_diff,e791:mixing,e791:cp0,e791:cp+,e791:dcsd}.
Throughout this paper, charge-conjugate modes are included unless 
otherwise noted.

The experimental apparatus consisted of a silicon vertex detector followed 
by a two-magnet spectrometer, two segmented Cerenkov counters for hadron 
identification, an electromagnetic calorimeter for electron identification, 
and iron shielding followed by scintillator counters for muon identification.
The silicon vertex detector consisted of 17 planes of silicon and was used to
reconstruct decay vertices downstream from the interaction vertex.
The spectrometer consisted of 35 planes of drift chambers
and two proportional wire chambers. The two dipole magnets bent 
particles in the horizontal plane and had \pt\ kicks of +212~GeV/$c$ 
and +320~GeV/$c$. The Cerenkov counters contained gases with different
indices of refraction; together they provided $\pi/K/p$ discrimination 
over the momentum range 6--60~GeV/$c$. More details about the detector
can be found in Ref.~\cite{e791:epj}.
 
Data was recorded using a loose transverse energy trigger. After 
reconstruction, events with evidence of well-separated interaction 
and decay vertices were retained for further analysis. Some of the
main criteria used to select charm decays are listed in Table~\ref{tab:cuts}.
The most important criterion is that of $SDZ$, defined as the distance 
between the interaction and decay vertices divided by the error 
in this quantity. Values used for this criterion ranged from 8 
(for $D^0$ and $D^+_s$ decays) to 20 (for longer-lived $D^+$ decays).

\begin{table}[htb]
\caption{Main criteria used to select $D$ decays.}
\label{tab:cuts}
\renewcommand{\arraystretch}{1.4}
\begin{tabular}{lc}
\hline
{\bf Selection criteria} & {\bf Typ. value} \\
\hline
$SDZ\ \equiv$ & \\
\hspace*{0.20in} $(z^{}_{\rm dec}-z^{}_{\rm int})/
\sqrt{\sigma^2_{\rm dec} + \sigma^2_{\rm int}}$ & 8--20 \\
\pt\ (transverse to $D$ direction) & $< 250$~MeV/$c$ \\
$min|z^{}_{\rm dec} - z^{}_{\rm target\ edge}|/\sigma^{}_{\rm sec}$ & $> 5$ \\
$D$ impact parameter &   \\
\hspace*{0.40in} w/r/t int. vertex & $< 60$~$\mu$m \\
$\chi^2_{\rm track}$ & $< 5$ \\
$t\ \equiv\ m^{}_D\,\times\,(z^{}_{\rm dec}-z^{}_{\rm int})/p$ & $< 5$~ps \\
\hline
\end{tabular}
\end{table}

\section{SEARCH FOR {\boldmath \ddbar} MIXING}

E791 has searched for \ddbar\ mixing via semileptonic 
$D^0\rightarrow K^+\ell^-\bar{\nu}$ decays and hadronic
$D^0\rightarrow K^+\pi^-$ and $D^0\rightarrow K^+\pi^-\pi^+\pi^-$ 
decays. For these searches we require that the $D^0$ be produced 
via $D^{*+}\rightarrow D^0\pi^+$ decay, and thus the flavor of 
the $D^0$ (or $\overline{D}^{\,0}$) when created is identified by 
the charge of the associated pion. The flavor of the $D^0$ when 
it decays is identified by the final state. With this information 
we measure the ratio $r^{}_{\rm mix}\equiv 
	\Gamma(D^0\rightarrow\overline{D}^{\,0}\rightarrow\bar{f})/
		\Gamma(D^0\rightarrow f)$. Each type of decay studied 
(semileptonic or hadronic) has an advantage and a disadvantage: 
the $K\ell\nu$ decays cannot be fully reconstructed due to
the missing neutrino, and thus the decay-time resolution is
degraded. The $K\pi/K\pi\pi\pi$ decays are fully reconstructed 
and thus have good time resolution, but they contain ``background'' 
arising from doubly-Cabibbo-suppressed (DCS) amplitudes that 
produce the same final state. The DCS amplitudes do not 
contribute to the semileptonic decays.

\subsection{Semileptonic {\boldmath $D^0\rightarrow K^+\ell^-\bar{\nu}$} Decays}
\vspace*{0.10in}

Semileptonic $D^0\rightarrow K^+\ell^-\bar{\nu}$ candidates were 
selected by requiring that there be a two-track vertex with $SDZ>8$.
One track was required to pass kaon identification criteria in the 
Cerenkov counters, and the other track was required to pass either 
electron identification criteria in the calorimeter or muon 
identification criteria in the scintillator counters following 
the iron shielding. Tracks identified as muons were required to have 
$p>10$~GeV/$c$ to reduce background from decays in flight. We define a 
quantity $M^{}_{\rm min}\equiv p^{}_T + \sqrt{p_T^2 + M^2_{K\ell}}$, 
where \pt\ is the transverse momentum of the $K\ell$ system with respect
to the $D^0$ direction-of-flight (obtained from the interaction and decay 
vertex positions), and $M^{}_{K\ell}$ is the invariant mass of the $K\ell$ 
pair. To reduce backgrounds, $M^{}_{\rm min}$  is required to be in the 
range 1.6--2.1 GeV/$c^2$, and $M^{}_{K\ell}$ is required to be in the 
range 1.15--1.80~GeV/$c^2$. The upper cut on $M^{}_{K\ell}$ removes 
background from $D^0\rightarrow K^-\pi^+$ decays in which the pion 
is misidentified as a lepton. After these cuts, the 
candidate $D^0$ is paired with a $\pi^\pm$ track originating from 
the interaction vertex (and having $p>2$~GeV/$c$) to form a $D^{*\pm}$. 

Since there is an undetected neutrino in the final state, the candidate
$D^0$ momentum cannot be measured directly. However, using the direction
of the $D^0$, the measured $K$ and $\ell$ momenta, and assuming the parent
particle mass to be that of a $D^0$, one can solve for the neutrino
momentum up to a two-fold ambiguity. We use the solution resulting
in higher $D^0$ momentum, as Monte Carlo (MC) studies indicate that
this provides a better estimate of the true momentum. From MC studies 
we determine that the r.m.s. difference between the calculated and 
the true $D^0$ momenta is about~15\%.

To search for a mixing signal, we divide the electron and muon 
samples into ``right-sign'' (RS) and ``wrong-sign'' (WS) decays. 
The former have the charge of the kaon being opposite to that 
of the pion from the $D^*$, whereas the latter have the charge 
of the kaon being the same as that of the pion. A WS decay 
would indicate \ddbar\ mixing. To determine the numbers of 
events in the four samples ($e$ and $\mu$, RS and WS), we 
perform an unbinned maximum likelihood fit using the $Q$ value 
($m^{}_{D^0\pi^+}-m^{}_{D^0}-m^{}_{\pi^+}$) and decay-time~$t$ 
($m^{}_{D^0}\times (z^{}_{\rm dec}-z^{}_{\rm int})/p$) for each 
event. For $D^{*+}\rightarrow D^0\pi^+$ decays, the $Q$ 
distribution is sharply peaked at 5.8~MeV.
We also include in the fits the $Q$ and $t$ distributions 
of background. The results for the electron sample are:
$N^{}_{RS}=1237\,\pm\,45$ and $N^{}_{WS}=4.4\,^{+11.8}_{-10.5}$. 
The results for the muon sample are:
$N^{}_{RS}=1267\,\pm\,44$ and $N^{}_{WS}=1.8\,^{+12.1}_{-11.0}$.
There is no indication of a mixing signal. Combining results 
from both $Ke\bar{\nu}$ and $K\mu\bar{\nu}$ samples gives
$r^{}_{\rm mix}=(0.11\,^{+0.30}_{-0.27})$\% or 
$r^{}_{\rm mix}<0.50\%$ at 90\% C.L. 

\subsection{Hadronic {\boldmath $D^0\rightarrow K^+\pi^-$ and
$D^0\rightarrow K^+\pi^-\pi^+\pi^-$} Decays}
\vspace*{0.10in}

\begin{figure}[htb]
\begin{center}
\mbox{\epsfig{file=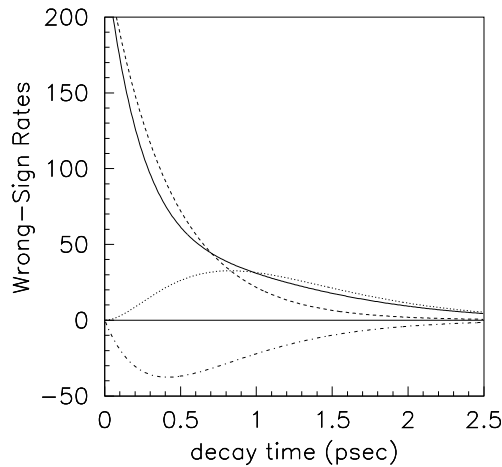,height=2.5in}}
\end{center}
\vspace*{-0.40in}
\caption{An example of the time dependence of $D^0\rightarrow K^+\pi^-$ 
decays due to the DCS amplitude (dashed), the mixing amplitude (dotted), 
and the interference between the two (dashed-dotted). The sum of
all three contributions is solid.}
\label{fig:timedep}
\end{figure}

$D^0\rightarrow K^+\pi^-$ candidates were selected from a sample of
two-prong decay vertices, and $D^0\rightarrow K^+\pi^-\pi^+\pi^-$
candidates were selected from four-prong vertices and from three-prong 
vertices with an extra track added. For the final event selection,
we use a two-layer neural network with 12 input variables (for $K\pi$) 
or 7 input variables (for $K\pi\pi\pi$). These variables include the
\pt\ of the $D^0$ candidate with respect to the $\pi^-$ beam direction, 
the \pt\ of the $D^0$ with respect to the $D^0$ direction (obtained
from the interaction and decay vertex positions), $SDZ$, the decay 
vertex fit $\chi^2$, the track fit $\chi^2$s, the Cerenkov counters' 
response for the $K$, etc. This selection results in eight
separate data sets: $D^0$ and $\overline{D}^{\,0}$, $K\pi$ and $K\pi\pi\pi$ 
final states, RS and WS decays. We subsequently perform a single unbinned 
maximum likelihood fit to all data sets, constructing the likelihood 
function from the kinetic energy $Q$, the decay-time $t$, and the reconstructed 
mass $m$ of each event. Backgrounds were carefully modeled and included 
in the fit. For each WS data set, the fit included contributions from 
a possible DCS amplitude. This amplitude results in a different $t$ 
dependence than that due to mixing, and this allows one to partially 
discriminate between the two sources. The full expression for the 
WS $t$ distribution (at small $t$ where we have acceptance) is:
\begin{eqnarray}
dN^{}_{\rm WS}/dt & \approx & e^{-\Gamma t}\ \times  
\label{eqn:timedep} \\
 &  & \hspace*{-0.50in} \Bigl(|{\cal A}^{}_{\rm DCS}|^2 + 
	|{\cal A}^{}_{\rm mix}|^2t^2 + 
		2Re({\cal A}^{}_{\rm DCS}{\cal A}^*_{\rm mix}) t\Bigr) .
\nonumber
\end{eqnarray}
An example of the time dependences of the three terms is 
plotted in Fig.~\ref{fig:timedep}.

The results of the fitting procedure depend upon whether we allow
interference between the DCS and mixing amplitudes, and also whether 
we allow \cp\ violation in any of the coefficients in Eq.~(\ref{eqn:timedep}).
In the most general case of allowing \cp\ violation in all
coefficients, we obtain
$r^{}_{\rm mix} (\overline{D}^{\,0}\rightarrow D^0) = 
			(0.18\,^{+0.43}_{-0.39}\,\pm 0.17)$\%
and 
$r^{}_{\rm mix} (D^0\rightarrow\overline{D}^{\,0}) = 
			(0.70\,^{+0.58}_{-0.53}\,\pm 0.18)$\%.
Allowing \cp\ violation in only the interference term gives
$r^{}_{\rm mix} = (0.39\,^{+0.36}_{-0.32}\,\pm 0.16)$\% or
$r^{}_{\rm mix} < 0.85$\% at 90\% C.L.

\begin{figure}[htb]
\begin{center}
\mbox{\epsfig{file=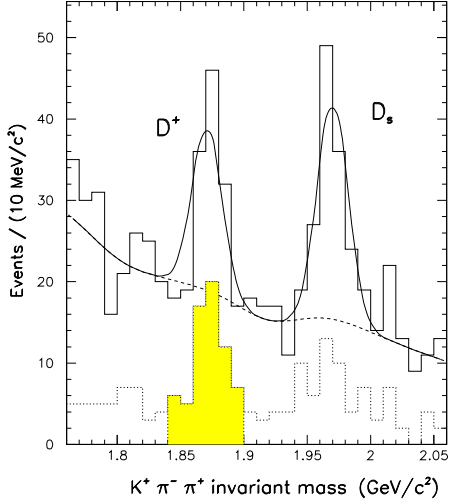,height=2.75in}}
\end{center}
\vspace*{-0.40in}
\caption{The $K^+\pi^-\pi^+$ invariant mass spectrum
for events passing final selection criteria. A peak 
resulting from the DCS decay $D^+\rightarrow K^+\pi^-\pi^+$ 
(and also from the singly-Cabibbo-suppressed decay 
$D^+_s\rightarrow K^+\pi^-\pi^+$) is clearly visible.}
\label{fig:dcsdplus}
\end{figure}

If we assume no mixing contribution (consistent with Standard Model
predictions at our level of sensitivity), we obtain a rate for DCS 
decays. The results are:
$r^{}_{\rm DCS}(K\pi) = (0.68\,^{+0.34}_{-0.33}\,\pm 0.07)$\% and
$r^{}_{\rm DCS}(K\pi\pi\pi) = (0.25\,^{+0.36}_{-0.34}\,\pm 0.03)$\%.
These values are approximately $\tan^4\theta^{}_C\times({\rm phase\ space})$, 
as expected, and are consistent with our measurement of the DCS charged 
decay $D^+\rightarrow K^+\pi^-\pi^+$ \cite{e791:dcsd}. For this latter 
measurement the final event sample is shown in Fig.~\ref{fig:dcsdplus}.
There are substantial backgrounds arising from misidentified charm decays
such as $D^0\rightarrow K^-\pi^+\pi^+$ and $D^+_s\rightarrow K^+K^-\pi^+$
(both Cabibbo-favored). We simultaneously fit for these backgrounds and
a $D^+\rightarrow K^+\pi^-\pi^+$ signal, finding $59\,\pm 13$ candidate 
events in the peak. This gives a measurement 
$r^{}_{\rm DCS}(K\pi\pi) = (0.77\,\pm\,0.17\,\pm\,0.08)$\%. 

\section{SEARCH FOR \cp\ VIOLATION}

E791 has searched for \cp\ violation in both charged and neutral
$D$ decays. For these searches we measure the time-integrated 
asymmetry $A^{}_{CP}$, defined as:
\begin{equation}
A^{}_{CP}\ \equiv\ \frac
	{\Gamma(D\rightarrow f) - \Gamma (\overline{D}\rightarrow\bar{f})}
	{\Gamma(D\rightarrow f) + \Gamma (\overline{D}\rightarrow\bar{f})}\ .
\end{equation}
We study only Cabibbo-suppressed final states, as for these
modes \cp-violating effects are expected to be largest. Because 
the incoming beam is~$\pi^-$, the production cross section for 
$\overline{D}^{\,0}$ and $D^-$ (in our acceptance) is a few percent 
larger than that for $D^0$ and $D^+$; this production asymmetry must 
be corrected for in order to discern a \cp\ asymmetry. To do this we 
define the ratios:
\begin{eqnarray}
\eta^{}_{D\rightarrow f} & \equiv & 
N^{}_{D\rightarrow f}/N^{}_{D\rightarrow K^-\pi^+(\pi^+)} 
\label{eqn:eta1} \\
\eta^{}_{\overline{D}\rightarrow\bar{f}} & \equiv &  
N^{}_{\overline{D}\rightarrow\bar{f}}/
		N^{}_{\overline{D}\rightarrow K^+\pi^-(\pi^-)}\ .
\label{eqn:eta2}  
\end{eqnarray}
Then $A^{}_{CP} = 
(\eta^{}_{D\rightarrow f} - \eta^{}_{\overline{D}\rightarrow\bar{f}})/
(\eta^{}_{D\rightarrow f} + \eta^{}_{\overline{D}\rightarrow\bar{f}})$
if
$\varepsilon^{}_{D\rightarrow f}/
		\varepsilon^{}_{D\rightarrow K^-\pi^+(\pi^+)} = 
\varepsilon^{}_{\overline{D}\rightarrow\bar{f}}/
		\varepsilon^{}_{\overline{D}\rightarrow K^+\pi^-(\pi^-)}$,
where $\varepsilon^{}_{D\rightarrow f}$ is the overall detection efficiency
for $D\rightarrow f$ (including acceptance). This relationship among 
efficiencies holds well for E791. We assume there is negligible 
\cp\ violation in the Cabibbo-favored decay modes used to normalize the 
Cabibbo-suppressed decay rates [Eqs.~(\ref{eqn:eta1}) and (\ref{eqn:eta2})].

\subsection{Neutral {\boldmath $D^0$} Decays}
\vspace*{0.10in}

\begin{figure}[htb]
\begin{center}
\hbox{
\hspace*{-0.50in}
\mbox{\epsfig{file=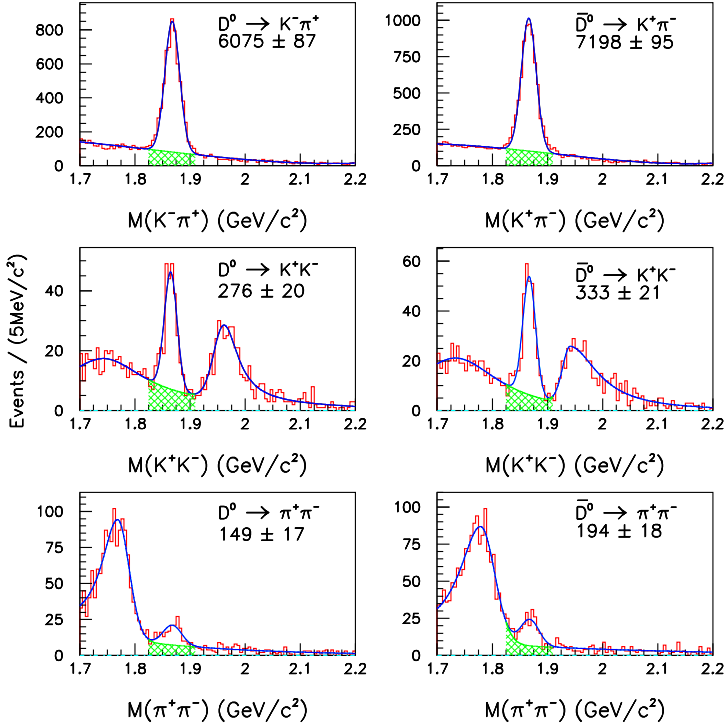,height=4.0in}}
}
\end{center}
\vspace*{-0.60in}
\caption{Final mass plots for $D^0$ (left) and
$\overline{D}^{\,0}$ (right) decays into final
states $K\pi$ (top row), $K^+K^-$ (middle 
row), and $\pi^+\pi^-$ (bottom row). }
\label{fig:kkpp}
\end{figure}

We measure $A^{}_{CP}$ for $D^0\rightarrow K^+K^-$ and
$D^0\rightarrow\pi^+\pi^-$, where the $D^0$ originates 
from $D^{*+}\rightarrow D^0\pi^+$ and thus the flavor 
of the $D^0$ is identified by the charge of the associated 
pion. The final mass plots are shown in 
Fig.~\ref{fig:kkpp} along with those for the normalization 
channel $D^0\rightarrow K^-\pi^+$. The solid curves superimposed
are our fits to the histograms. The integrals of the Gaussian 
distributions used for the signals determine the number of 
$D^0\rightarrow K\pi/KK/\pi\pi$ decays. These event yields 
(listed in Fig.~\ref{fig:kkpp}) are used to calculate 
$\eta^{}_{D^0}$ and $\eta^{}_{\overline{D}^{\,0}}$ and 
subsequently determine $A^{}_{CP}$.  The results are:
\begin{eqnarray}
A^{}_{CP}(K^+ K^-) & = & 
	\hspace*{-0.075in} -0.010\,\pm\,0.049\,\pm\,0.012 \hspace*{0.40in}  \\
A^{}_{CP}(\pi^+\pi^-) & = &
	\hspace*{-0.075in} -0.049\,\pm\,0.078\,\pm\,0.030\,. 
\end{eqnarray}
These correspond to 90\% confidence intervals
$-9.3\% < A^{}_{CP}(K^+K^-) < 7.3\%$ and 
$-18.6\% < A^{}_{CP}(\pi^+\pi^-) < 8.8\%$. 

\subsection{Charged {\boldmath $D^+$} Decays}
\vspace*{0.10in}

\begin{figure}
\begin{center}
\mbox{\epsfig{file=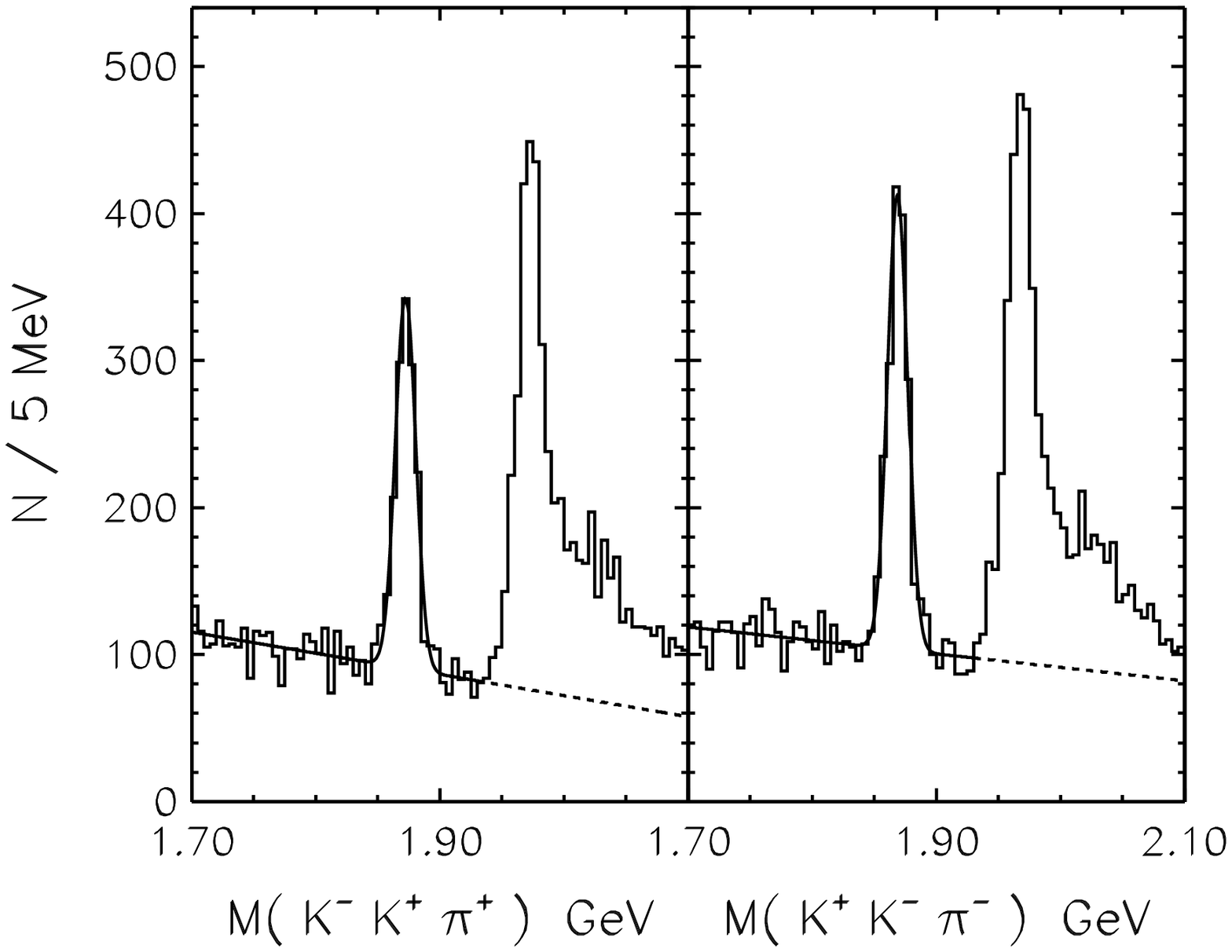,height=1.8in}}
\vspace*{0.02in}
\mbox{\epsfig{file=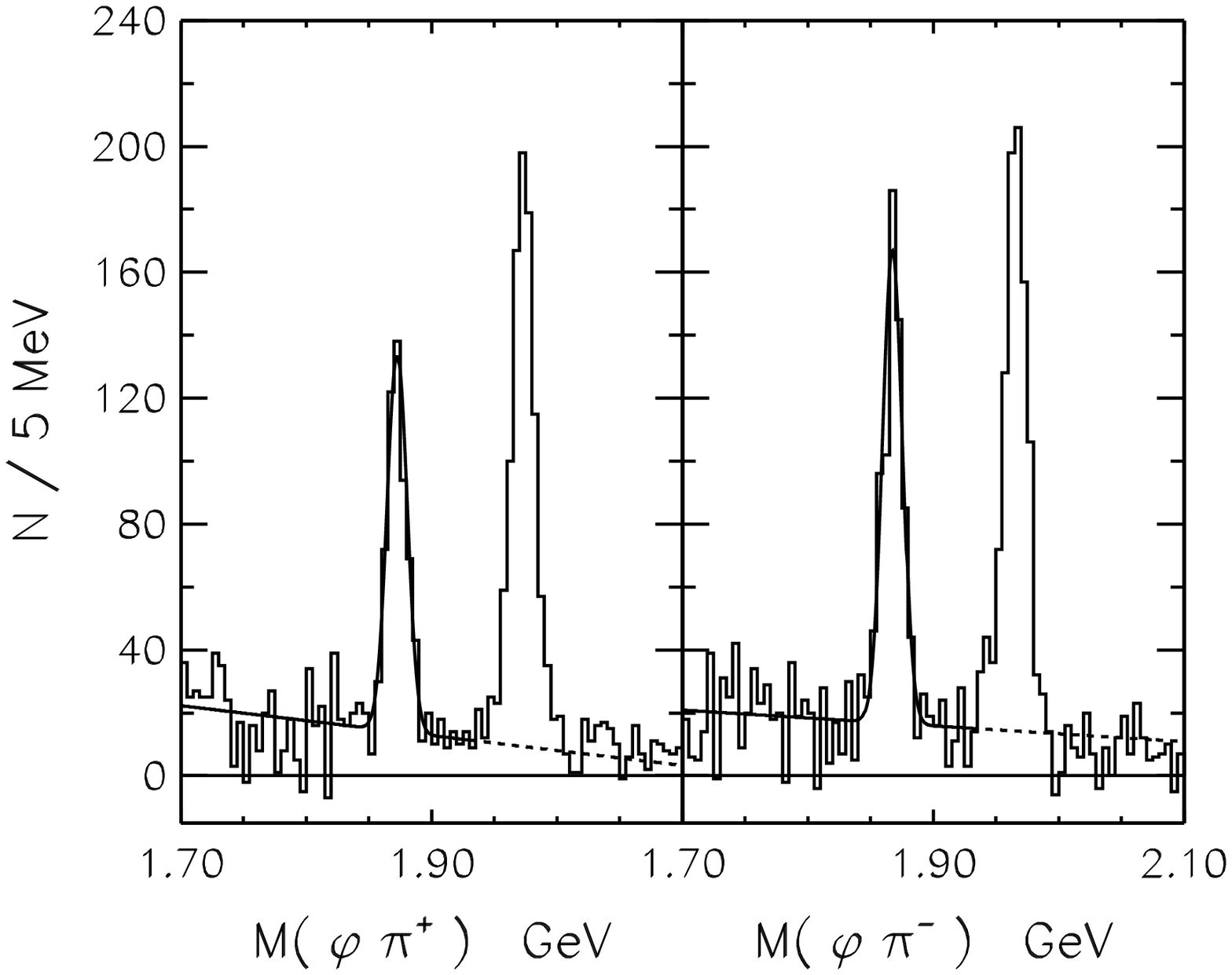,height=1.8in}}
\vspace*{0.02in}
\mbox{\epsfig{file=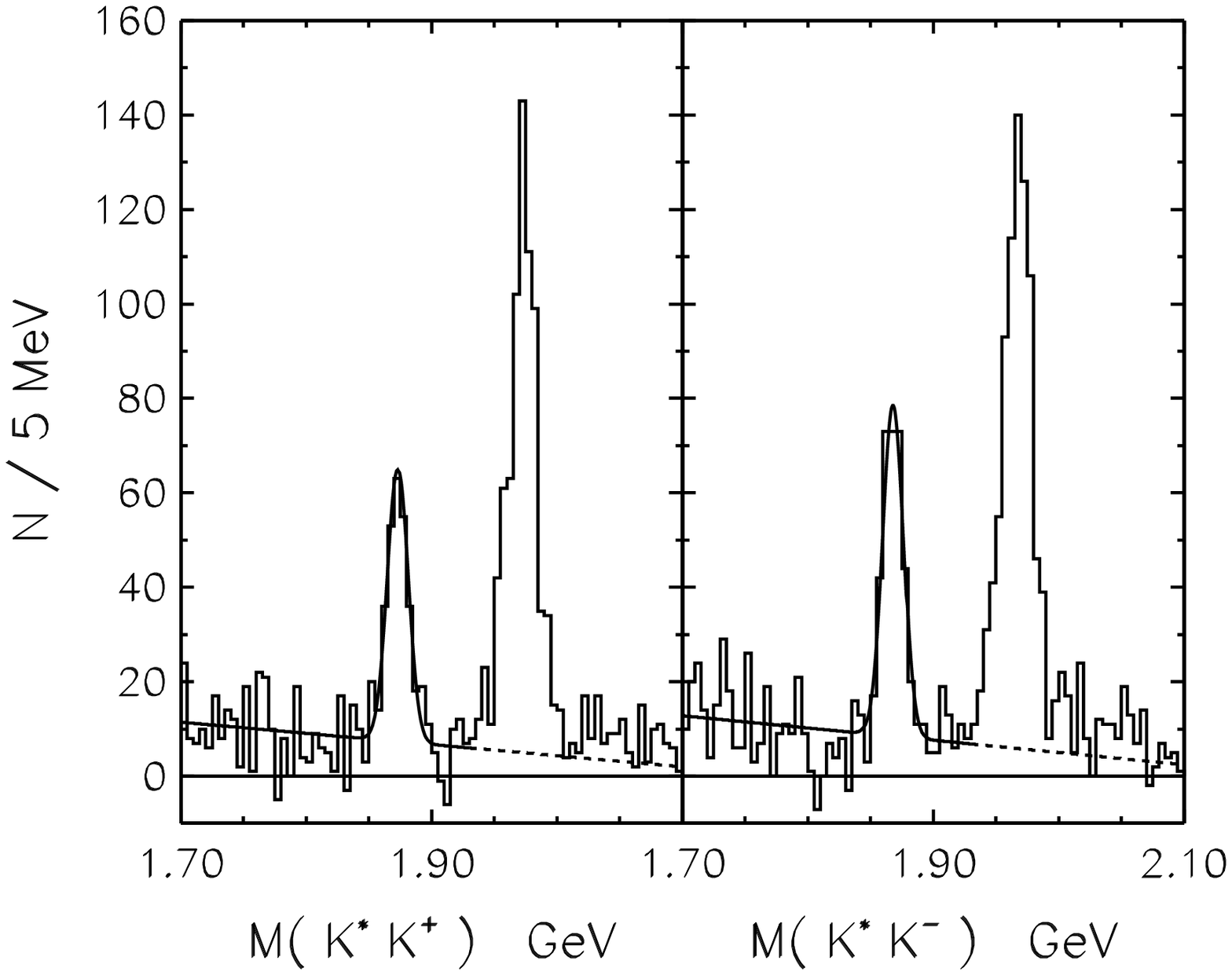,height=1.8in}}
\vspace*{0.02in}
\mbox{\epsfig{file=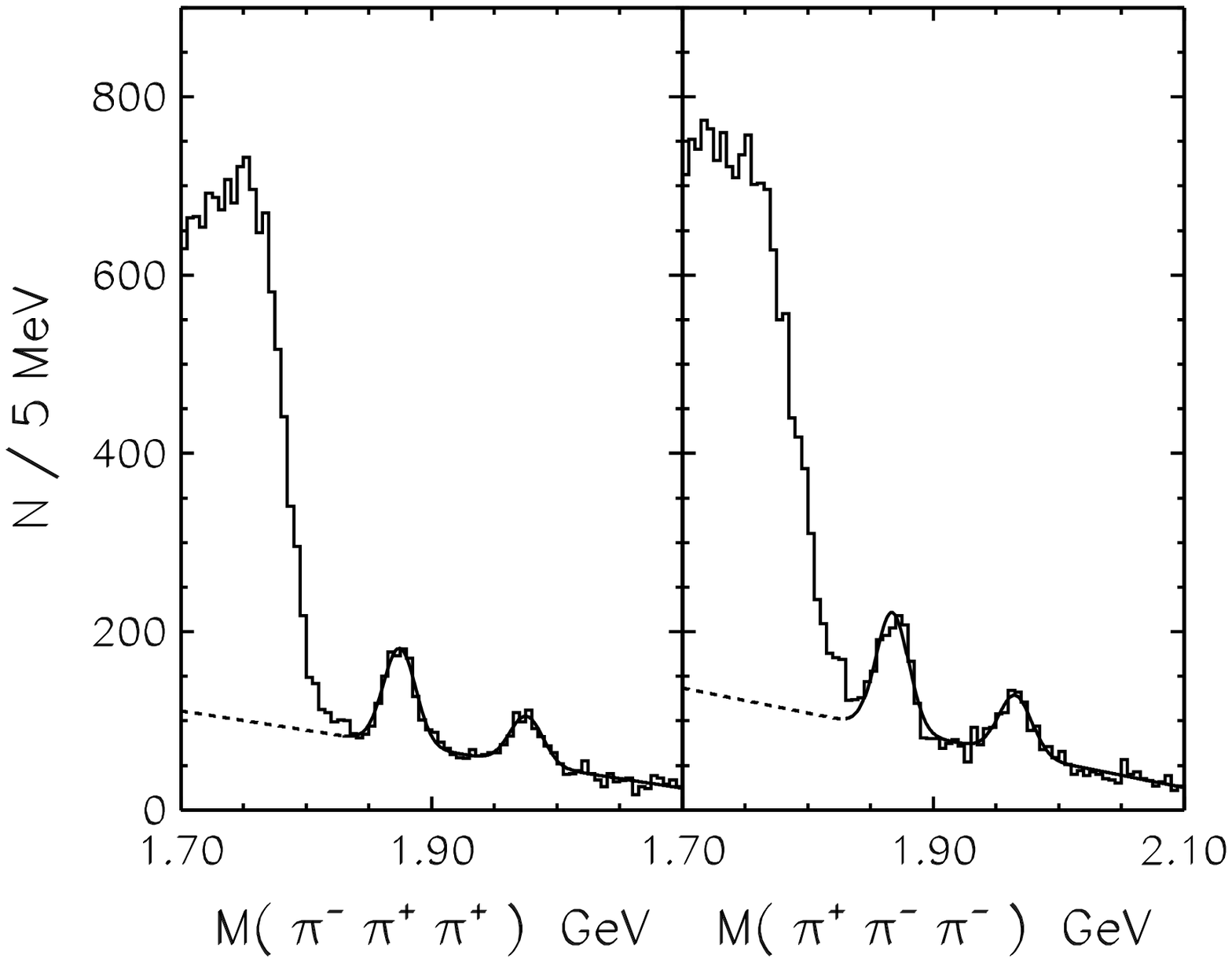,height=1.8in}}
\end{center}
\vspace*{-0.40in}
\caption{Final mass plots for $D^+$ (left) and $D^-$ (right)
decays into final states $\phi\,\pi$, $\overline{K}^{*0}K$, 
$KK\pi$ (nonresonant), and $\pi\pi\pi$.}
\label{fig:d+cp}
\end{figure}

We measure $A^{}_{CP}$ for $D^+$ decays into the final states 
$\phi\,\pi^+\ (\phi\rightarrow K^+K^-)$, 
$\overline{K}^{*0}K^+\ (\overline{K}^{*0}\rightarrow K^-\pi^+)$, 
$K^+K^-\pi^+$ (nonresonant), and $\pi^+\pi^-\pi^+$. For all modes, 
the normalization channel is $D^+\rightarrow K^-\pi^+\pi^+$. For 
the $\phi\,\pi^+$ final state we require  
$|m^{}_{K^+K^-}-m^{}_\phi | < 6$~MeV/$c^2$;\ \,for the 
$\overline{K}^{*0}K^+$ final state we require
$|m^{}_{K^-\pi^+}-m^{}_{K^*} | < 45$~MeV/$c^2$. 
The mass spectra for the final event samples are shown 
in Fig.~\ref{fig:d+cp}, and the measured asymmetries 
$A^{}_{CP}$ are listed in Table~\ref{tab:d+cp}. We also 
list measurements from other experiments, and for each decay 
mode we calculate a 90\% confidence interval for $A^{}_{CP}$ 
incorporating all measurements listed. As the measurements 
are from independent experiments, we assume their statistical
and systematic errors uncorrelated.
We observe that $A^{}_{CP}$ for $D^+\rightarrow K^+K^-\pi^+$ 
is relatively well-constrained:
the 90\% CL interval is $-1.6$\% to $+2.0$\%. 
For $D^0\rightarrow\pi^+\pi^-\pi^+$, the E791 
measurement is the only result available.

\begin{table*}[htb]
\caption{\cp\ asymmetries measured for $D^0$ and $D^+$ decays. 
Also listed are measurements from other experiments, and listed in 
boldface type are 90\% confidence intervals incorporating all 
the experimental results.}
\label{tab:d+cp}
\renewcommand{\arraystretch}{1.4}
\begin{tabular}{l|cc}
\hline
{\bf Mode} &  {\bf {\boldmath $A^{}_{CP}$ (E791)}}
	   &  {\bf {\boldmath $A^{}_{CP}$ (Others)}}  \\
\hline
$D^0\rightarrow K^+ K^-$  &  $-0.010\,\pm\,0.049\,\pm\,0.012$
     &  \begin{tabular}{c}
    $+0.024\,\pm\,0.084$ (E687 \cite{e687:cp}) \\ 
    $+0.080\,\pm\,0.061$ (CLEO \cite{cleo}) \\ 
    $-0.001\,\pm\,0.022\,\pm\,0.015$ (E831 \cite{e831:cp}) 
        \end{tabular} \\
 &  \multicolumn{2}{c}{\bf {\boldmath $-2.6<A^{}_{CP}<4.4$ \%}\hspace*{0.50in}} \\
\hline
$D^0\rightarrow \pi^+ \pi^-$  &  $-0.049\,\pm\,0.078\,\pm\,0.030$
     &  $+0.048\,\pm\,0.039\,\pm\,0.025$ (E831 \cite{e831:cp}) \\ 
 &  \multicolumn{2}{c}{\bf {\boldmath $-4.1<A^{}_{CP}<9.2$ \%}\hspace*{0.50in}} \\
\hline
$D^+\rightarrow K^+ K^- \pi^+$  &  $-0.014\,\pm\,0.029$
     &  \begin{tabular}{c}
    $-0.031\,\pm\,0.068$ (E687 \cite{e687:cp}) \\ 
    $+0.006\,\pm\,0.011\,\pm\,0.005$ (E831 \cite{e831:cp}) 
        \end{tabular} \\
 &  \multicolumn{2}{c}{\bf {\boldmath $-1.6<A^{}_{CP}<2.0$ \%}\hspace*{0.50in}} \\
\hline
$D^+\rightarrow \phi\,\pi^+$  &  $-0.028\,\pm\,0.036$
			     & $0.066\,\pm\,0.086$ (E687 \cite{e687:cp})  \\
 &  \multicolumn{2}{c}{\bf {\boldmath $-6.8<A^{}_{CP}<4.1$ \%}\hspace*{0.50in}} \\
\hline
$D^+\rightarrow\overline{K}^{*0}K^+$  &  $-0.010\,\pm\,0.050$
			&  $-0.12\,\pm\,0.13$ (E687 \cite{e687:cp}) \\
 &  \multicolumn{2}{c}{\bf {\boldmath $-10<A^{}_{CP}<5.2$ \%}\hspace*{0.50in}} \\
\hline
$D^+\rightarrow \pi^+\pi^-\pi^+$  &  $-0.017\,\pm\,0.042$
     &  \\
 &  \multicolumn{2}{c}{\bf {\boldmath $-8.6<A^{}_{CP}<5.2$ \%}\hspace*{0.50in}} \\
\hline
\end{tabular}
\end{table*}

\section{MEASUREMENT OF THE WIDTH DIFFERENCE {\boldmath $\Delta\Gamma$}}

E791 has measured the difference in decay widths between the two
mass-eigenstates of the $D^0$-$\overline{D}^{\,0}$ system. This provides
a measurement of the mixing parameter $y\equiv \Delta\Gamma/(2\overline{\Gamma})$.
Theoretically, 
$r^{}_{\rm mix} = \Gamma(D^0\rightarrow\overline{D}^{\,0}\rightarrow\bar{f})/
\Gamma(D^0\rightarrow f) = (x^2 + y^2)/2$ for $x,y$ small, where
$x\equiv \Delta m/\overline{\Gamma}$.

The method used to determine $\Delta\Gamma$ is as follows:
assuming no \cp\ violation, the two mass-eigenstates are
\cp\ eigenstates and can be written
$D^{}_1=(|D^0\rangle + |\overline{D}^{\,0}\rangle )/\sqrt{2}$ and
$D^{}_2=(|D^0\rangle - |\overline{D}^{\,0}\rangle )/\sqrt{2}$.
Observing a $K^+K^-$ final state ($CP=+1$) denotes a 
$D^{}_1\rightarrow K^+K^-$ decay, and $dN^{}_{KK}/dt\propto e^{-\Gamma^{}_1 t}$.
Observing a $K^-\pi^+$ or $K^+\pi^-$ final state denotes a 
$D^0$ or $\overline{D}^{\,0}$ decay, respectively (neglecting DCS 
amplitudes for simplicity). In this case 
$dN^{}_{K\pi}/dt\propto e^{-\overline{\Gamma} t}\cosh (\Delta\Gamma/2)t$
\cite{ajs:tdistribution},
where $\overline{\Gamma}= (\Gamma^{}_1 + \Gamma^{}_2)/2$ 
and $\Delta\Gamma = \Gamma^{}_1-\Gamma^{}_2$. Our previous 
limit $r^{}_{\rm mix}<0.50\%$ implies $|y| < 0.10$ or 
$|\Delta\Gamma | < 0.48$~ps$^{-1}$, and thus 
$\cosh (\Delta\Gamma/2)t\approx 1$ for the range 
of lifetime $t$ accepted by the experiment. Thus 
$dN^{}_{K\pi}/dt\propto e^{-\overline{\Gamma} t}$, and
$\Delta\Gamma = 2\left(\Gamma^{}_{KK} - \Gamma^{}_{K\pi}\right)$.
Equivalently, $y = \Delta\Gamma/(2\overline{\Gamma}) = 
			\tau^{}_{K\pi}/\tau^{}_{KK} - 1$.
From an experimental point of view, it is convenient that 
$\cosh (\Delta\Gamma/2)t\approx 1$
as we actually measure $t-t^{}_{\rm cut}\equiv t'$, and while
$e^{-\overline{\Gamma}t'}\propto e^{-\overline{\Gamma}t}$,
$\cosh (\Delta\Gamma/2)t'$ is {\it not\/} proportional to 
$\cosh (\Delta\Gamma/2)t$.

\begin{figure}[htb]
\begin{center}
\mbox{\epsfig{file=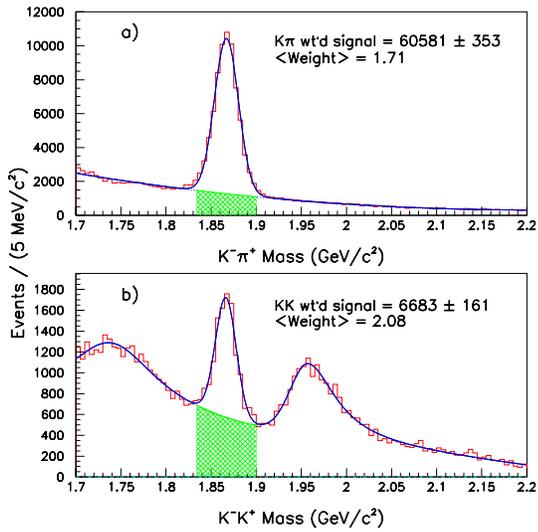,height=2.85in}}
\end{center}
\vspace*{-0.40in}
\caption{Final mass plots for $D^0\rightarrow K^-\pi^+$ (top) 
and $D^0\rightarrow K^+K^-$ (bottom). The peak to the right of 
the $D^0\rightarrow K^+K^-$ peak (bottom plot) is due to 
$D^0\rightarrow K^-\pi^+$ decays in which the pion 
has been misidentified as a kaon.}
\label{fig:lifediff1}
\end{figure}

\begin{figure}[htb]
\begin{center}
\mbox{\epsfig{file=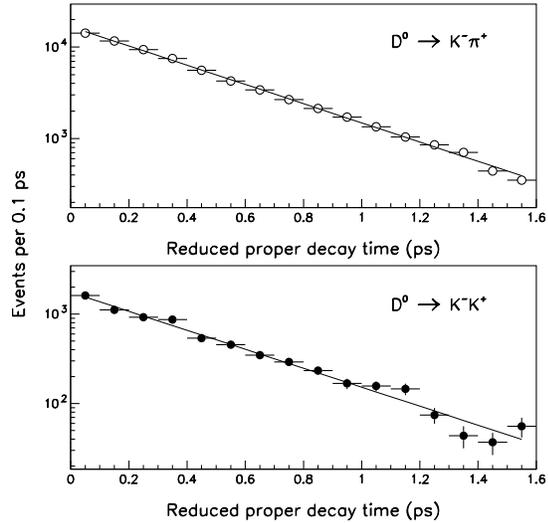,height=2.85in}}
\end{center}
\vspace*{-0.40in}
\caption{Reduced lifetime ($t'$) distributions 
for $D^0\rightarrow K^-\pi^+$ (top) and
$D^0\rightarrow K^+K^-$ (bottom). The 
contributions from background have been subtracted.}
\label{fig:lifediff3}
\end{figure}

The $D^0\rightarrow K^-\pi^+$ and $D^0\rightarrow K^+K^-$ final
event samples are shown in Fig.~\ref{fig:lifediff1}. The resultant 
lifetime distributions (for $t'$) are shown in Fig.~\ref{fig:lifediff3}. 
These distributions have the background shape subtracted, and fitting to 
them yields:
$\Gamma^{}_{K\pi}=2.420\,\pm\,0.019$~ps$^{-1}$ and 
$\Gamma^{}_{KK}=2.441\,\pm\,0.068$~ps$^{-1}$. 
The difference in widths $\Delta\Gamma$ is 
$0.04\,\pm\,0.14\,\pm\,0.05$~ps$^{-1}$, where the first error 
is statistical (resulting from the fits) and the second error 
is the sum in quadrature of the systematic errors. These 
systematic errors are listed in Table~\ref{tab:syserrors}. 
The result corresponds to a 90\% confidence interval 
$-0.20<\Delta\Gamma < 0.28$~ps$^{-1}$. 
For $y$ we obtain $0.008\,\pm 0.029\,\pm\,0.010$ or 
$-0.042 < y < 0.058$ at 90\% C.L. Combining this with 
a recent measurement by FNAL E831 
($y=0.0342\,\pm 0.0139\,\pm\,0.0074$ \cite{e831:gamma_diff}) 
gives $0.006 < y < 0.052$ at 90\% C.L.

\begin{table}[htb]
\caption{Systematic errors contributing to the measurement of $\Delta\Gamma$.}
\label{tab:syserrors}
\renewcommand{\arraystretch}{1.4}
\begin{tabular}{l|ccc}
\hline
 & {\bf {\boldmath $\tau^{}_{K\pi}$ }} & 
{\bf {\boldmath $\tau^{}_{KK}$ }} & 
{\bf {\boldmath $\Delta\Gamma$ }} \\
{\bf {\boldmath Systematic error}}  & 
	{\bf (ps)} & {\bf (ps)} & {\bf {\boldmath (ps$^{-1}$)}} \\ 
\hline
Fit range 		& 0.002 & 0.003 & 0.024 \\
Selection criteria	& 0.001 & 0.002 & 0.020 \\
Particle ID weighting	& 0.001 & 0.003 & 0.024 \\
MC production model	& 0.003 & 0.003 & 0.000 \\
Fixed width		& 0.001 & 0.002 & 0.030 \\
\hline
Total & 0.004 & 0.006 & 0.050 \\
\hline
\end{tabular}
\end{table}

\section{SUMMARY}
We have searched for \ddbar\ mixing in both semileptonic and hadronic
$D^0$ decays and see no evidence for it. We subsequently set 90\% CL upper 
limits $r^{}_{\rm mix}$\,$<$\,0.50\% and $r^{}_{\rm mix}$\,$<$\,0.85\%, 
respectively. Assuming the mixing amplitude to be negligibly small 
(as predicted by the Standard Model), we fit our data for the rate 
of DCS decays $D^0\rightarrow K^+\pi^-$ and 
$D^0\rightarrow K^+\pi^-\pi^+\pi^-$, obtaining
$r^{}_{\rm DCS}(K\pi) = (0.68\,^{+0.34}_{-0.33}\,\pm 0.07)$\% and
$r^{}_{\rm DCS}(K\pi\pi\pi) = (0.25\,^{+0.36}_{-0.34}\,\pm 0.03)$\%.
These results are consistent with the rate we measure for
the DCS charged decay $D^+\rightarrow K^+\pi^-\pi^+$:
$r^{}_{\rm DCS}(K\pi\pi) = (0.77\,\pm\,0.17\,\pm\,0.08)$\%. 

We have also searched for \cp\ violation in neutral
$D^0\rightarrow K^+K^-/\pi^+\pi^-$ decays and in charged
$D^+\rightarrow \phi\,\pi^+/\overline{K}^{*0}K^+/K^+K^-\pi^+/\pi^+\pi^-\pi^+$ 
decays. We see no evidence for \cp\ violation, and for
each mode we set 90\% C.L. limits on the asymmetry
parameter $A^{}_{CP}$. These results are listed in
Table~\ref{tab:d+cp}. 

Finally, we measure the difference in decay widths 
$\Delta\Gamma$ between the two mass-eigenstates of the 
\ddbar\ system. We convert this into a measurement of the 
mixing parameter $y = \Delta\Gamma/(2\overline{\Gamma})$.
Our result is $y=0.008\,\pm 0.029\,\pm\,0.010$ or 
$-0.042 < y < 0.058$ at 90\% C.L. Combining this
with a recent measurement by FNAL E831 
gives $0.006 < y < 0.052$ at 90\% C.L.

\end{document}